\documentstyle[12pt]{article}
\begin{document}
\input{FEYNMAN}
\def
\dslash{\partial \!\!\! /}
\def
\dbar{\bar\Delta}

%
%
\def\ap#1#2#3{     {\it Ann. Phys. (NY) }{\bf #1} (19#2) #3}
\def\arnps#1#2#3{  {\it Ann. Rev. Nucl. Part. Sci. }{\bf #1} (19#2) #3}
\def\ijmpa#1#2#3{  {\it Int. J. Mod. Phys. }{\bf A#1} (19#2) #3}
\def\npb#1#2#3{    {\it Nucl. Phys. }{\bf B#1} (19#2) #3}
\def\plb#1#2#3{    {\it Phys. Lett. }{\bf B#1} (19#2) #3}
\def\prd#1#2#3{    {\it Phys. Rev. }{\bf D#1} (19#2) #3}
\def\prb#1#2#3{    {\it Phys. Rev. }{\bf B#1} (19#2) #3}
\def\pro#1#2#3{    {\it Phys. Rev. }{\bf #1} (19#2) #3}
\def\prep#1#2#3{   {\it Phys. Rep. }{\bf #1} (19#2) #3}
\def\pha#1#2#3{    {\it Physica }{\bf A#1} (19#2) #3}
\def\prl#1#2#3{    {\it Phys. Rev. Lett. }{\bf #1} (19#2) #3}
\def\ptp#1#2#3{    {\it Prog. Theor. Phys. }{\bf #1} (19#2) #3}
\def\rmp#1#2#3{    {\it Rev. Mod. Phys. }{\bf #1} (19#2) #3}
\def\zpc#1#2#3{    {\it Z. Physik }{\bf C#1} (19#2) #3}
\def\mpla#1#2#3{   {\it Mod. Phys. Lett. }{\bf A#1} (19#2) #3}
\def\sjnp#1#2#3{   {\it Sov. J. Nucl. Phys. }{\bf #1} (19#2) #3}
\def\yf#1#2#3{     {\it Yad. Fiz. }{\bf #1} (19#2) #3}
\def\nc#1#2#3{     {\it Nuovo Cim. }{\bf #1} (19#2) #3}
\def\jetp#1#2#3{   {\it Sov. Phys. JETP }{\bf #1} (19#2) #3}
\def\jetpl#1#2#3{  {\it JETP Lett. }{\bf #1} (19#2) #3}
\def\ibid#1#2#3{   {\it ibid. }{\bf #1} (19#2) #3}
\def\ijmpa#1#2#3{  {\it Int. J. Mod. Phys. }{\bf A#1} (19#2) #3}
\def\el#1#2#3{     {\it Europhys. Lett. }{\bf #1} (19#2) #3}
\def\jmmm#1#2#3{   {\it J. Magn. Magn. Mater. }{\bf #1} (19#2) #3}
\def\jpa#1#2#3{     {\it J. Phys. }{\bf A#1} (19#2) #3}

April 15, 1995\hspace{7cm} IP-ASTP-10195
\vspace{50pt}
\begin{center}
{\large\sc{\bf Decimation in more then one dimension.}}

\baselineskip=12pt
\vspace{50pt}

V. Kushnir* and B. Rosenstein**,

\vspace{20pt}
Institute of Physics, Academia Sinica\,
Taipei, 11529\,
Taiwan, R.O.C.\,

\vspace{50pt}
\end{center}
\begin{abstract}
We develop a formalism for
performing real space renormalization group transformations
of the "decimation type"
using perturbation theory. The type of transformations beyond
$d=1$ is nontrivial even for free theories. We check the formalism on
solvable case of $O(N)$ symmetric Heisenberg chain.
 The transformation is particularly useful
to study asymptotically free theories. Results for one class
of such models, the d=2 O(N) symmetric
$\sigma$ models ($N\ge 3$)
for decimation with scale factor $\eta=2$ (when quarter of the points is left)
are given as an example.
\end{abstract}
\vspace{30pt}

*PHVK@PHYS.SINICA.EDU.TW \\

**BARUCH@PHYS.SINICA.EDU.TW

\newpage
The renormalization group (RG) transformations is one of the most powerful
and frequently used conceptual as well as practical tools
in statistical physics and quantum field theory.
While conceptually the idea of
combining variables on neighbouring sites into complexes is
very simple, in practise it almost always turns out to be rather
complicated.

Historically, usefulness of RG transformations was
realized after the d=2 Ising model on triangular lattice
was very elegantly solved by Niemeijer and Van Leeuwen \cite{Niemeijer-Leeuwen}
using block spinning.
The nonlinear RG method they used was peculiar to that particular system
and did not allow generalization to more complicated cases.
For more general systems Wilson \cite{Wilson}
proposed to use the weak coupling perturbation theory in momentum space. This
was first applied to
scalar $\phi^4$ models and subsequently to spin systems \cite{Nelson}
and lattice gauge theories for thinning by factor 2 \cite{Mutter}.
For complicated systems like these with local gauge symmetries
some approximate methods were
 developed like Migdal-Kadanoff \cite{Migdal-Kadanoff}
approximation, variational RG \cite{Patkos}, mean-field RG \cite{Fittipaldi}
or block spinning using Monte Carlo numerical methods
\cite{Swendsen,Shenker-Tobochnik}.
However, unlike perturbation theory, these approximations are
uncontrollable in a sense that it is not clear how to estimate errors.

Exact RG transformations are generally not known
(exceptions are decimations in spin chains $d=1$ and mentioned above very
special cases in $d=2$). Moreover after one RG transformation
the resulting action contains generically infinite number of
interaction terms, and therefore one is forced to make an additional
approximation
dropping some of them (hopefully the less relevant ones).
RG transformations are especially useful when applied repeatedly.
This requires self similarity of the approximate effective action
and is justified only around fixed points. Note however that accurate
thinning of the lattice even just by factor $\eta=2$ can greatly facilitate
the study of a model by means of subsequent MC simulation.

In this paper we perform the decimation for multidimensional free theories and
then develop a systematic formalism to calculate perturbatively a decimated
effective action for interacting models. \footnote{Method, based on the Schur
formula, somewhat similar to ours, was developed for quasiperiodic systems on
the octagonal lattice \cite{Bellissard}.}

There are several types of the RG transformations. The conceptually
simplest one
is the decimation or thinning of degrees of freedom in the configuration space.
Some degrees of freedom located, for example, on sites with at least one
odd coordinate are simply integrated out.

\begin{picture}(40000,30000)(0,-14000)
\drawline\fermion[\N\REG](2000,-5000)[19000]
\drawline\fermion[\N\REG](7000,-5000)[19000]
\drawline\fermion[\N\REG](12000,-5000)[19000]
\drawline\fermion[\N\REG](17000,-5000)[19000]
\drawline\fermion[\N\REG](22000,-5000)[19000]
\drawline\fermion[\E\REG](0,-3000)[24000]
\drawline\fermion[\E\REG](0,2000)[24000]
\drawline\fermion[\E\REG](0,7000)[24000]
\drawline\fermion[\E\REG](0,12000)[24000]
\put(2000,-3000){\circle*{1000}}
\put(12000,-3000){\circle*{1000}}
\put(22000,-3000){\circle*{1000}}
\put(7000,2000){\circle{1000}}
\put(17000,2000){\circle{1000}}
\put(2000,7000){\circle*{1000}}
\put(12000,7000){\circle*{1000}}
\put(22000,7000){\circle*{1000}}
\put(7000,12000){\circle{1000}}
\put(17000,12000){\circle{1000}}
\put(7000,-3000){\circle{1000}}
\put(17000,-3000){\circle{1000}}
\put(2000,2000){\circle{1000}}
\put(12000,2000){\circle{1000}}
\put(22000,2000){\circle{1000}}
\put(7000,7000){\circle{1000}}
\put(17000,7000){\circle{1000}}
\put(2000,12000){\circle{1000}}
\put(12000,12000){\circle{1000}}
\put(22000,12000){\circle{1000}}

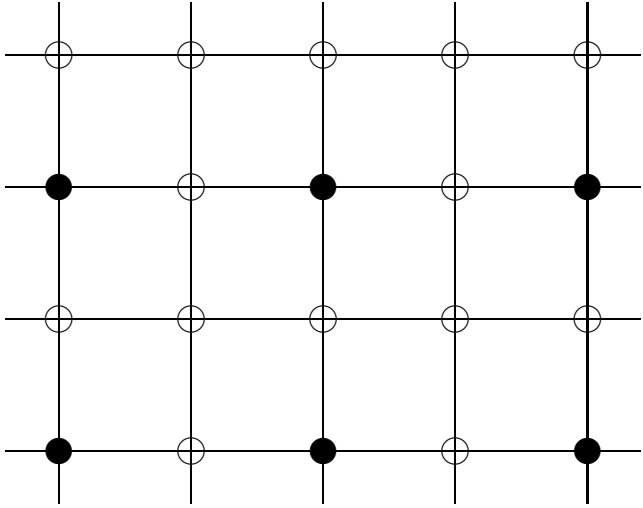
\put(-2000,-9000){Fig. 1. Decimation: full circles belong to sublattice ${\cal
L}^*$,}
 \put(-2000,-10500){ while empty circles $x \in {\cal D}={\cal L}-{\cal L}^*$
denote the integrated out sites.}

\end{picture}

 Example is given
on Fig. 1 on which spins at empty circle points are integrated out.
\begin{equation}
Z=\sum_{\phi(X):X \in {\cal L}^*}\sum_{\phi(x):x \in {\cal L}-{\cal
L}^*}e^{-{\cal A}[\phi(x),\phi(X)]}=\sum_{\phi(X):X \in {\cal L}^*}e^{-{\cal
A}^{dec}[\phi(X)]}
\label{Z}
\end{equation}
Here and in what follows points of the coarse lattice are denoted by capital
letters.
The resulting effective action $A^{dec}$ contains generally
interactions of any range. Here ${\cal L}$ is the original $d$ dimensional
lattice (lattice spacing $a$), while ${\cal L}^*$ is a sublattice with lattice
spacing $A=\eta a$. In what follows $A$ will set a scale: $A=1$. Note that
remaining
variables are all the old variables. This is not the case for
the so called block spin transformations.
One defines a linear or a nonlinear combination of the variables
on ${\cal L}$, the block spin:
\begin{equation}
{\bf \phi}(X)=f[X,\phi(x)]
\label{<O>}
\end{equation}
For example, for the $O(N)$ classical spins $S^a$ one can define
\cite{Shenker-Tobochnik}
\begin{equation}
{\bf S}^a(X)=\sum_{block X} S^a(x)/|\sum_{block X} S^a(x)|
\label{block}
\end{equation}
The choice of the combination is highly ambiguous and success of the
transformation
critically depends on it. The main
problem is that it is extremely
difficult in practise to perform such a transformation even
perturbatively.
The relations like eq.(\ref{block}) are very nonlinear and even singular
\cite{Griffiths}.

Another type of RG transformations, used especially extensively
in field theory, is the momentum space RG \cite{Wilson,Polchinski}.
One defines the momentum space variables
\begin{equation}
\phi(p)\equiv 1/(2 \pi)^d \int d^d x  e^{i p x} \phi(x)
\label{f(p)}
\end{equation}
Now one performs integration over high frequence modes
(strictly speaking
chopping the Brillouin zone, but more often the approximate spherically
symmetric momentum cutoff $\Lambda$ is utilized \cite{Nelson,Ma}).
This type of RG transformations, while convenient for the $\phi^4$ model, turns
out to be especially inconvenient
for constrained systems like the $O(N)$ symmetric $\sigma$ model.
The reason is following.
While generally in x - space the constraint are local, for example
\begin{equation}
S^a(x) S^a(x)=1
\label{constraint}
\end{equation}
in $p$ - space it becomes a convolution. What does it mean now
high frequency physical modes? The constraint mixes between low and
high frequencies. Since most systems of interest belong to this
class one has to circumvent the difficulty.
One way is to solve the constraint and make the momentum space RG
for physical quantities only. Then
the mode integrated
effective action contains generally "noncovariant terms". The original
global symmetry is lost since the high frequency modes
do not constitute an $O(N)$ symmetric set. Problems are more
acute with local gauge symmetries.
In practice this type of thinning out of degrees of freedom is often
used for the demonstration purposes only and very rarely the actual
calculations.

Decimation are extremely difficult
to perform even in free theory in more then one dimension (see,
for example \cite{Hu}). It might sound surprising that something
is difficult in free theory since all the integrals are gaussian
and "doable in principle".
Of course it is still a gaussian integral, but a very complicated one.
Let us consider a free massless boson nearest neighbours action
\begin{equation}
{\cal A}[\phi]=-\frac {a^{(d-2)}}{2}\sum_{xy}\phi(x)\Box(x-y)\phi(y)
\label{f}
\end{equation}
where the lattice Laplacian is defined by
$\Box(x)\equiv
\sum_{\mu=1}^d \left[\delta(x-\mu)+\delta(x+\mu)-2\delta(x)\right]$.

If one tries to integrate out a point $\phi(0,...,0)$, the gaussian integral
involves all its $2d$ nearest neighbours.
$$\int d \phi(0)\exp\left[ -\frac{a^{(d-2)}}{2}
\left( 2 d \phi(0)^2 -2 \phi(0)\sum_\mu(\phi(\mu)+\phi(-\mu) \right) \right]=
$$
\begin{equation}
=\exp\left[\frac {a^{(d-2)}}{2d} \left(\sum_\mu
\phi(\mu)^2+\sum_{\mu\ne\nu}\phi(\mu)\phi(\nu)\right)\right]
\end{equation}
This is very simple.
However when trying subsequently to
integrate another point, say $(2,0...,0)$, all the previous point's neighbours
enter the gaussian integral and so on.
The gaussian integration requires inverting
increasingly larger
matrices. Since we have to integrate out all the points not belonging to
the sublattice, some other methods are required. An exception is
the $d=1$ case.
Here the size of the matrix does not grow: integration of a point leads
just to interactions of the neighbouring remaining points.
This is the reason why it is possible in many cases to explicitly find
decimations in $d=1$.

Let us return with free massless boson on lattice eq.(\ref{f}). Effective
action,
after decimation with parameter $\eta$ generally has a form:
\begin{equation}
{\cal A}^A[\phi(X)]=-\frac{1}{2}\sum_{X\in R^d}\phi(X){\bf \Delta}(X-Y)\phi(Y)
\end{equation}
where bold letters denote sublattice functions.

Using corresponding Fourier transforms we write propagators on the lattice and
the sublattice:
$$
G(x)=\frac {1}{(2\pi)^d}\int_{-\pi/a}^{\pi/a} d^dk e^{- i k x a} \frac {a^2}{4
\sum_\mu {\rm sin}^2 (ak_\mu /2)}
$$
\begin{equation}
{\bf G}(X)=\frac {1}{2\pi}\int_{-\pi}^{\pi}e^{- i K X}\frac {1}{{\bf
\Delta}(K)}
\end{equation}

The two actions should give an equivalent correlators between
two sublattice points: 0 and $X$:
$G(\eta X)={\bf G}(X)$.
This leads to the following relation between the Fourier transforms:
\begin{equation}
{\bf G}(K)=1/\Delta(K)= \frac {1}{(2\pi)^d}\sum_X e^{-i K X}
\int_{-\pi/a}^{\pi/a} d^dk e^{i k X}
\frac {a^2}{\sum_\mu 4\; {\rm sin}^2 (ak_\mu/2)}
\end{equation}
Summation over $X$ results in sum over $\delta$ functions
\begin{equation}
{\bf G}(K)= \frac {a^d}{(2\pi)^{d}}\int_{-\pi/a}^{\pi/a} d^dk
\sum_{n=-\infty}^\infty \delta ((k-K+2\pi n)_\mu)
\frac {a^2}{\sum_\mu 4\; {\rm sin}^2 (ak_\mu/2)}
\end{equation}
which are used to perform the integrations over $k$:
\begin{equation}
G(K)=
\sum_{n_\mu=1}^{\eta}\frac {a^2}{\sum_\mu 4\; {\rm sin}^2 (a(K+2\pi n)_\mu/2)}
\label{decprop}
\end{equation}
Limits of summation follow from the different sizes of Brillouin
 zone for two lattices.
In $d=1$ we recover the original form of the action, while for $d>1$, the sum
over one of the variables, $n_1$ can be explicitly done \cite{Prudnikov},
but the remaining summations should be done numerically.
In particular for $d=2$ the propagator is:
\begin{equation}
{\bf G}(K)=\frac{1}{\eta}\sum_{n_2 = 1}^{\eta}
   \frac { {\rm sinh} (\eta \, \alpha) {\rm csch} (\alpha ) }
      {2 \left( -1 +
           {\rm \cosh} (\eta \, \alpha) + 2 {\rm sin}^2 (\frac {K_1}{2})
\right) },
\end{equation}
where
$$
\alpha={\rm arccosh}(1 + 2{\rm sin}^2 (\frac {K_2 + 2\pi n_2}{2\eta})).
$$
For large $\eta$ the euclidean invariance is restored,
 ${\bf G}(K_1,K_2)={\bf G}({\sqrt {K_1^2+K_2^2}})$
 and
numerical calculations show that it can be fitted by
\begin{equation}
\label{fit}
G(K)=\frac {1}{K^2} + \frac{1}{2 \pi}\log(\eta)+0.04876 + 0.003022\, K^2 +
O[(K^2)^2]
\end{equation}
 with an accuracy better 1 percent in all the Brillouin zone.
 Note that the decimated propagator even for large $\eta$
does not coincides with the naive continuum limit $1/K^2$.
The contact constant term with logarithmic dependence of $\eta$
 is typical for $d=2$ and is nothing else but the bubble integral.
The polynomial coefficients are very small and almost coincide with
the Loran expansion of the propagator around $K=0$.
For finite $\eta$ the symmetry remains of course just the discrete subgroup
of the rotations.
In higher dimensions similar expressions can be written.
The procedure can be extended also to free fermion fields.

Now we  briefly outline a systematic procedure for
decimation-type RG transformations in interacting models.
 For concreteness we discuss the lattice $\phi^4$
model
\begin{equation}
{\cal A}[\phi_x]=a^{(d-2)} \sum_x \left[ \frac{1}{2}(\nabla \phi_x)^2+\frac{m^2
a^2}{2}\phi^2_x
+\frac{\lambda a^2}{4!}\phi^4_x \right],
\end{equation}
where $\nabla \phi_x= (\phi_x -\phi_{x-1} )$.

In the process of decimation, the fields $\phi_X$ on the sublattice ($X\in
{\cal L}^*$) will be
treated as "external sources".

\begin{picture}(40000,27000)(0,-14000)
\put(500,8000){\circle*{1000}}
\put(500,6000){a}
\drawline\fermion[\E\REG](10000,8000)[6000]
\put(13000,6000){b}
\put(9500,8000){\circle*{1000}}
\put(16500,8000){\circle{1000}}
\drawline\fermion[\E\REG](23000,8000)[6000]
\put(22500,8000){\circle{1000}}
\put(29500,8000){\circle{1000}}
\put(26000,6000){c}
\drawline\fermion[\E\REG](3500,-1000)[3000]
\drawline\fermion[\E\REG](7500,-1000)[3000]
\drawline\fermion[\N\REG](7000,-4500)[3000]
\drawline\fermion[\N\REG](7000,-500)[3000]
\put(7000,-1000){\circle*{1000}}
\put(7000,-6000){d}
\drawline\fermion[\E\REG](18500,-1000)[3000]
\drawline\fermion[\E\REG](22500,-1000)[3000]
\drawline\fermion[\N\REG](22000,-4500)[3000]
\drawline\fermion[\N\REG](22000,-500)[3000]
\put(22000,-1000){\circle{1000}}
\put(22000,-6000){e}

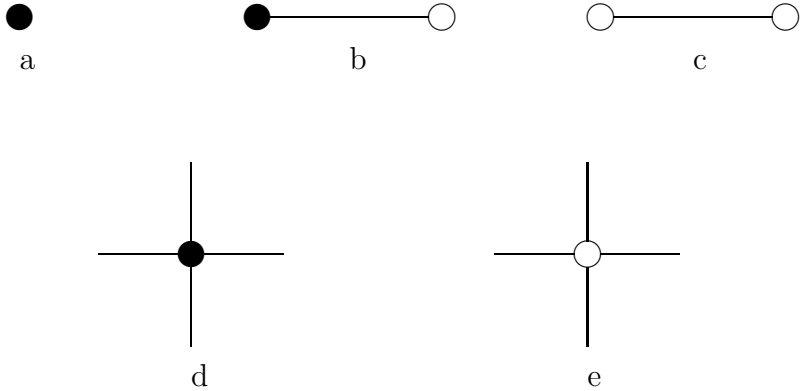
\put(0,-8000){Fig.2. Real space RG (decimation) propagators
 (a, b, c) and vertices (d, e)}
\put(0,-9500){for $\phi^4$ model. Full circles belong to sublattice (external
fields),}
\put(0,-11000){while empty denote "internal fields" ($x\in {\cal L}-{\cal
L}^*$).}

\end{picture}

 Action can be divided into three parts: "classical action" of
the external sources $\phi_X$ (Fig. 2a, 2d),
$$
{\cal A}^{cl}[\phi_X]=\sum_X\left( \frac {a^{d} m^2}{2} \phi^2_X+ a^{d-2} d\;
\phi^2_X+ \frac {a^{d} \lambda}{4!}\phi^4_X\right)
$$
\begin{equation}
\equiv \frac {a^{(d-2)}}{2}\sum_{X,Y} \phi_X A_{XY}\phi_Y+A^{cl}_{int}[\phi_X],
\label{ext}
\end{equation}
cross-term (Fig. 2b)
\begin{equation}
{\cal A}_1[\phi_x,\phi_X]=-a^{(d-2)}\sum_{x ,X} \phi_X \dbar(X,x) \phi_x\equiv
-a^{(d-2)}\sum_{x ,X} \phi_x B_{x X} \phi_X
\end{equation}
with "external legs"
\begin{equation}
\dbar(X,x)= \sum_\mu(\delta_{X-x+\mu}+\delta_{X-x-\mu})
\label{cross}
\end{equation}
 and an internal part for which all the vertices belong to ${\cal D}\equiv
 {\cal L}-{\cal L}^*$ ("decorated" model) (Fig. 2c, 2e):
\begin{equation}
{\cal A}_2[\phi_x]=-\frac {a^{(d-2)}}{2} \sum_{x,y} \phi_x D(x,y) \phi_y+\frac
{a^{d} \lambda}{4!}\sum_x \phi^4_x.
\label{int}
\end{equation}
 Note that, unlike momentum space RG, here external fields $\phi_X$ are coupled
to internal part only via derivative couplings like off-diagonal part of
propagator (Fig. 2b) or derivative interaction in nonlinear $\sigma$-model. All
the local vertices will completely decouple into internal (Fig.2e) and external
(Fig. 2d).

Integration out all the fields $\phi_x$ will lead to an effective action for
the fields $\phi_X$ on sublattice of the form:
\begin{equation}
{\cal A}[\phi_X]=\sum_{X_1,..,X_{2n}} \frac {1}{(2 n)!}
H^{(2n)}(X_1,..,X_{2n})\phi_{X_1}...\phi_{X_{2n}},
\label{acteff}
\end{equation}
where the coefficient functions $H^{(2n)}(X_1,..,X_{2n})$
 are sums of all the connected contributions with $2n$ ends.
These connected functions does not degenerate into one particle irreducible.

To integrate over field $\phi_x$ perturbatively, we need to find its propagator
which is matrix inverse to $D$. At the same time, all we explicitly have is the
full Laplacian $\Delta$ and original propagator $G=\Delta^{-1}$. Moreover,
since ${\cal D}$ does not constitute a sublattice, it is impossible
to make use of Fourier analysis on it. Therefore it is useful to
represent all the summations over ${\cal D}$ via summations over
${\cal L}$ and ${\cal L}^*$.
This can be done using following algebraic trick. Matrix $\Delta$ as well
 as matrix $G$ can be decomposed into following blocks
\begin{equation}
\Delta=\left( \matrix{A&B\cr
                  B^t&D} \right),
\end{equation}
\begin{equation}
G=\left( \matrix{a&b\cr
                  b^t&d} \right),
\end{equation}
Infinite-dimensional matrices $A, B, D$ defined by the quadratic part of action
eqs. (\ref{ext}, \ref{cross}, \ref{int}) (Fig. 2a, 2b, 2c). The matrices $b,d$
are the usual propagator matrices (Fig. 2b, 2c)  and $a$ is the propagator
between the points of sublattice, that is an expression inverse to $G(X)$. Now
we can invert $D$ using the fact that matrices $\Delta$ and $G$ are inverse to
each other. This implies a set of algebraic relations for their submatrices:
$$
A a + B b^t=1,\,
A b + B d=0,\,
B^t a + D b^t=0,\,
B^t b + D d=1
$$
and, after straightforward transformations we obtain an expression for the
"internal" propagator:
\begin{equation}
D^{-1}=d-b^t a b
\end{equation}
or, returning to previous notations,
\begin{equation}
\label{int-prop}
D^{-1}_{xy}=G_{xy}-\sum_{X,Y}G_{xX} {\bf G}^{-1}_{XY} G_{Yy}
\end{equation}
\begin{picture}(10000,3000)(0,-2000)
\put(5000,-300){$=$}
\drawline\fermion[\E\REG](7000,0)[5000]
\put(13000,-300){$-$}
\drawline\fermion[\E\REG](16000,0)[5000]
\THICKLINES
\drawline\fermion[\E\REG](\pbackx,0)[5000]
\THINLINES
\put(\fermionfrontx,0){\circle{500}}
\drawline\fermion[\E\REG](\pbackx,0)[5000]
\put(\fermionfrontx,0){\circle{500}}
\end{picture}

Here and in what follows bold line denotes free inverse decimated propagator
${\bf G}^{-1}_{XY}$, while thin line corresponds to propagator of the original
theory $G_{xy}$.
Using this representation of $D^{-1}_{xy}$ as an internal line, we can now
begin to build the perturbation theory. We would like to stress that using of
$D^{-1}_{xy}$ eq.(\ref{int-prop}) enables us to extend this summation over $\it
all$ the original lattice. Indeed one can see that this expression is equal to
zero when at least one of the points $x, y$ in $D^{-1}_{xy}$ belongs to ${\cal
L}^*$

Calculation of an n-point function $H^{(n)}(X_1,..,X_n)$ in effective action
eq.(\ref{acteff}) for the coarse-grained field $\phi_X$ will be as follows.

 All connected diagrams with $n$ end points $X_1,..,X_n$, $V$ vertices and $I$
internal lines in real space are drawn with following components:

a) all vertices are situated at the points $x\in {\cal L}$ and to every vertex
at point $x_i$ corresponds summation $\sum_{x_i}$;

b) $\dbar_{Xx}$ are assigned to external ends $X$ and

c) the internal lines $D^{-1}_{xy}$ are represented via eq.(\ref{int-prop}).
This representation splits each diagram into $2^I$ subdiagrams and each of
these subdiagrams should be calculated separately. This calculation includes
summation over all the internal points $x_i$ on the fine grained lattice and
over the internal sublattice points $Y$ [end points of the inverse decimated
propagator ${\bf G}^{-1}$, see eq.(\ref{int-prop})]. Finally the "classical
contribution" to coefficient function should be added.

For the sake of simplicity let us discuss calculations of the decimation
diagrams on the concrete example. Namely, we will consider one of the
contributions to four point function $H^{(4)}$ in $\phi^4$ model (Fig. 3).

With using of the propagator $D^{-1}_{xy}$, original diagram splits into five
subdiagrams (Fig. 3a, 3b, 3c, 3d, 3e). Typical subdiagram here (for instance,
subdiagram d) can be written as
$$
H^{(4)}_d (X_1,..,X_4)=
$$
\begin{equation}
\sum\limits_{\scriptstyle y_1,..,y_4,z \atop \scriptstyle Y_1,..Y_4}
\dbar_{X_1\; y_1} G_{y_1\; Y_1} {\bf G}^{-1}_{Y_1\; Y_2} G_{Y_2\; z}
\dbar_{X_2\; y_2} G_{y_2\; Y_3} {\bf G}^{-1}_{Y_3\; Y_4} G_{Y_4\; z}
\dbar_{X_3\; y_3} G_{x_3\; z} \dbar_{X_4\; y_4} G_{x_4\; z}.
\label{fourp}
\end{equation}
As usual, in practical calculations it is very convenient to employ the Fourier
transformed functions at intermediate steps. In this way we will operate with
vertices, legs
\begin{equation}
\dbar(k)=2 a^{d-2} \sum_\mu {\rm cos}(k_\mu a),
\end{equation}
propagators $G(k)$ and inverse decimated propagator ${\bf G}^{-1}(K)$
eq.(\ref{decprop}). However, it is not convenient to perform Fourier transform
of expression eq.(\ref{fourp}) immediately, because it contains functions
defined on different lattices.

\begin{picture}(40000,48000)(0,-34000)
\drawline\fermion[\SE\REG](0,10000)[9000]
\drawline\photon[\NW\REG](\fermionfrontx,\fermionfronty)[2]
\put(\pbackx,\pbacky){\circle{500}}
\drawline\photon[\SE\REG](\fermionbackx,\fermionbacky)[2]
\put(\pbackx,\pbacky){\circle{500}}
\drawline\fermion[\NE\REG](0,\fermionbacky)[9000]
\global\Yone=\pmidy
\drawline\photon[\SW\REG](\fermionfrontx,\fermionfronty)[2]
\put(\pbackx,\pbacky){\circle{500}}
\drawline\photon[\NE\REG](\fermionbackx,\fermionbacky)[2]
\put(\pbackx,\pbacky){\circle{500}}

\global\advance \Yone by -300
\global\advance \fermionfrontx by 11000
\put(\fermionfrontx,\Yone){$\longrightarrow$}

\drawline\fermion[\SE\REG](20000,10000)[9000]
\drawline\photon[\NW\REG](\fermionfrontx,\fermionfronty)[2]
\put(\pbackx,\pbacky){\circle{500}}
\drawline\photon[\SE\REG](\fermionbackx,\fermionbacky)[2]
\put(\pbackx,\pbacky){\circle{500}}
\drawline\fermion[\NE\REG](20000,\fermionbacky)[9000]
\drawline\photon[\SW\REG](\fermionfrontx,\fermionfronty)[2]
\put(\pbackx,\pbacky){\circle{500}}
\drawline\photon[\NE\REG](\fermionbackx,\fermionbacky)[2]
\put(\pbackx,\pbacky){\circle{500}}
\put(22500,0){(a)}

\put(0,-7500){$-$}

\put(2000,-7500){$\left ( \right.$}

\drawline\fermion[\SE\REG](5000,-4000)[1500]
\global\Ytwo=\fermionbacky
\drawline\photon[\NW\REG](\fermionfrontx,\fermionfronty)[2]
\put(\pbackx,\pbacky){\circle{500}}
\THICKLINES
\drawline\fermion[\SE\REG](\fermionbackx,\fermionbacky)[1500]
\THINLINES
\put(\pfrontx,\pfronty){\circle{500}}
\drawline\fermion[\SE\REG](\pbackx,\pbacky)[6000]
\put(\fermionfrontx,\fermionfronty){\circle{500}}
\drawline\photon[\SE\REG](\fermionbackx,\fermionbacky)[2]
\put(\pbackx,\pbacky){\circle{500}}
\drawline\fermion[\NE\REG](5000,\fermionbacky)[9000]
\drawline\photon[\SW\REG](\fermionfrontx,\fermionfronty)[2]
\put(\pbackx,\pbacky){\circle{500}}
\drawline\photon[\NE\REG](\fermionbackx,\fermionbacky)[2]
\put(\pbackx,\pbacky){\circle{500}}
\put(7500,-14000){(b)}

\global\advance \fermionbackx by 1000
\put(\fermionbackx,-7500){$+$ ... $\left. \right )$}

\put(17500,-7500){$+$}

\put(20000,-7500){$\left ( \right.$}

\drawline\fermion[\SE\REG](23000,-4000)[1500]
\drawline\photon[\NW\REG](\fermionfrontx,\fermionfronty)[2]
\put(\pbackx,\pbacky){\circle{500}}
\THICKLINES
\drawline\fermion[\SE\REG](\fermionbackx,\fermionbacky)[1500]
\THINLINES
\put(\pfrontx,\pfronty){\circle{500}}
\drawline\fermion[\SE\REG](\pbackx,\pbacky)[6000]
\put(\pfrontx,\pfronty){\circle{500}}
\drawline\photon[\SE\REG](\fermionbackx,\fermionbacky)[2]
\put(\pbackx,\pbacky){\circle{500}}
\drawline\fermion[\NE\REG](23000,\fermionbacky)[1500]
\drawline\photon[\SW\REG](\fermionfrontx,\fermionfronty)[2]
\put(\pbackx,\pbacky){\circle{500}}
\THICKLINES
\drawline\fermion[\NE\REG](\fermionbackx,\fermionbacky)[1500]
\THINLINES
\put(\pfrontx,\pfronty){\circle{500}}
\drawline\fermion[\NE\REG](\pbackx,\pbacky)[6000]
\put(\pfrontx,\pfronty){\circle{500}}
\drawline\photon[\NE\REG](\fermionbackx,\fermionbacky)[2]
\put(\pbackx,\pbacky){\circle{500}}
\put(25000,-14000){(c)}

\put(30000,-7500){$+$ ... $\left. \right )$}

\put(0,-21500){$-$}

\put(2000,-21500){$\left( \right. $}

\drawline\fermion[\SE\REG](5000,-18000)[1500]
\drawline\photon[\NW\REG](\fermionfrontx,\fermionfronty)[2]
\put(\pbackx,\pbacky){\circle{500}}
\THICKLINES
\drawline\fermion[\SE\REG](\fermionbackx,\fermionbacky)[1500]
\THINLINES
\put(\pfrontx,\pfronty){\circle{500}}
\drawline\fermion[\SE\REG](\pbackx,\pbacky)[3000]
\put(\pfrontx,\pfronty){\circle{500}}

\THICKLINES
\drawline\fermion[\SE\REG](\fermionbackx,\fermionbacky)[1500]
\THINLINES
\put(\pfrontx,\pfronty){\circle{500}}
\drawline\fermion[\SE\REG](\pbackx,\pbacky)[1500]
\put(\pfrontx,\pfronty){\circle{500}}

\drawline\photon[\SE\REG](\fermionbackx,\fermionbacky)[2]
\put(\pbackx,\pbacky){\circle{500}}
\drawline\fermion[\NE\REG](5000,\photonfronty)[1500]
\drawline\photon[\SW\REG](\fermionfrontx,\fermionfronty)[2]
\put(\pbackx,\pbacky){\circle{500}}
\THICKLINES
\drawline\fermion[\NE\REG](\fermionbackx,\fermionbacky)[1500]
\THINLINES
\put(\pfrontx,\pfronty){\circle{500}}
\drawline\fermion[\NE\REG](\pbackx,\pbacky)[6000]
\put(\pfrontx,\pfronty){\circle{500}}

\drawline\photon[\NE\REG](\fermionbackx,\fermionbacky)[2]
\put(\pbackx,\pbacky){\circle{500}}
\put(7500,-28000){(d)}

\global\advance \fermionbackx by 1000
\put(\fermionbackx,-21500){$+$ ... $\left. \right )$}

\put(18000,-21500){$+$}

\drawline\fermion[\SE\REG](23000,-18000)[1500]
\drawline\photon[\NW\REG](\fermionfrontx,\fermionfronty)[2]
\put(\pbackx,\pbacky){\circle{500}}
\THICKLINES
\drawline\fermion[\SE\REG](\fermionbackx,\fermionbacky)[1500]
\THINLINES
\put(\pfrontx,\pfronty){\circle{500}}
\drawline\fermion[\SE\REG](\pbackx,\pbacky)[3000]
\put(\pfrontx,\pfronty){\circle{500}}

\THICKLINES
\drawline\fermion[\SE\REG](\fermionbackx,\fermionbacky)[1500]
\THINLINES
\put(\pfrontx,\pfronty){\circle{500}}
\drawline\fermion[\SE\REG](\pbackx,\pbacky)[1500]
\put(\pfrontx,\pfronty){\circle{500}}

\drawline\photon[\SE\REG](\fermionbackx,\fermionbacky)[2]
\put(\pbackx,\pbacky){\circle{500}}
\drawline\fermion[\NE\REG](23000,\photonfronty)[1500]
\drawline\photon[\SW\REG](\fermionfrontx,\fermionfronty)[2]
\put(\pbackx,\pbacky){\circle{500}}
\THICKLINES
\drawline\fermion[\NE\REG](\fermionbackx,\fermionbacky)[1500]
\THINLINES
\put(\pfrontx,\pfronty){\circle{500}}
\drawline\fermion[\NE\REG](\pbackx,\pbacky)[3000]
\put(\pfrontx,\pfronty){\circle{500}}
\THICKLINES
\drawline\fermion[\NE\REG](\fermionbackx,\fermionbacky)[1500]
\THINLINES
\put(\pfrontx,\pfronty){\circle{500}}
\drawline\fermion[\NE\REG](\pbackx,\pbacky)[1500]
\put(\pfrontx,\pfronty){\circle{500}}

\drawline\photon[\NE\REG](\fermionbackx,\fermionbacky)[2]
\put(\pbackx,\pbacky){\circle{500}}
\put(25000,-28000){(e)}
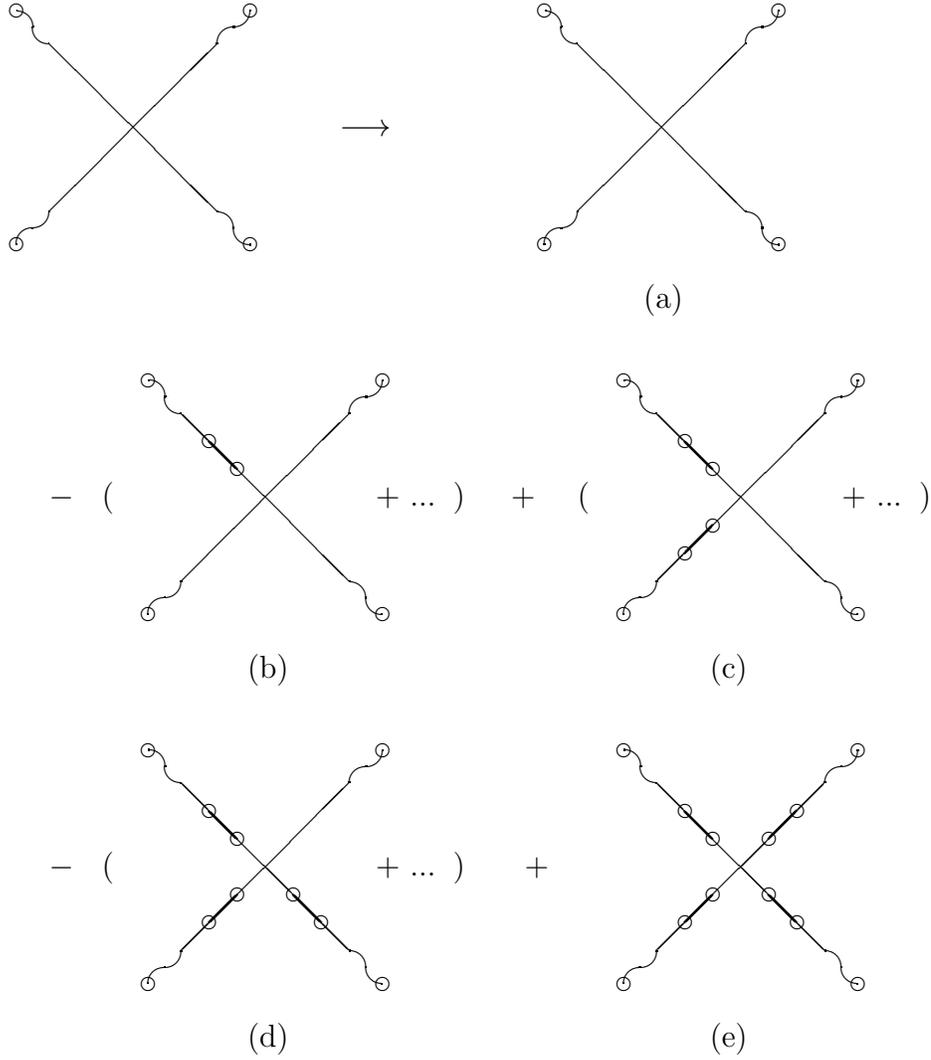
\put(0,-31000){Fig.3. Four point contribution to the $\phi^4$ effective action}

\end{picture}

 Instead, we can use the fact that this expression breaks into blocks where all
internal points lie on the fine grained lattice ${\cal L}$ and sublattice
points enter only as ends. These blocks are connected by ${\bf
G}^{-1}(Y_i,Y_j)$. In terms of such an "${\cal L}$-connected" parts the diagram
eq.(\ref{fourp}) has the form:
\begin{equation}
\label{blockdiagr}
H^{(4)}_d (X_1,..,X_4)=\sum\limits_{ \scriptstyle Y_1,..Y_4} F_1(X_1,\; Y_1)
F_2(X_2,\; Y_2) F_3(X_3,\;X_4,\;Y_3,\; Y_4){\bf G}^{-1}_{Y_1\; Y_2}{\bf
G}^{-1}_{Y_3\; Y_4}
\end{equation}
with blocks
\begin{equation}
F_1(X_1,\; Y_1)=\sum\limits_{y_1} \dbar_{X_1\; y_1} G_{y_1\; Y_1},
\end{equation}
\begin{equation}
F_2(X_2,\; Y_2)= \sum\limits_{y_2} \dbar_{X_2\; y_2} G_{y_2\; Y_3}
\end{equation}
and
\begin{equation}
F_3(X_3,\;X_4,\;Y_3,\; Y_4)=\sum\limits_{y_3,y_4,z} G_{Y_2\; z} G_{Y_4\; z}
\dbar_{X_3\; y_3} G_{x_3\; z} \dbar_{X_4\; y_4} G_{x_4\; z}.
\end{equation}
The most economic way to proceed is to Fourier transform each block in
eq.(\ref{blockdiagr}) with respect to the fine grained lattice variables only.
This will result in diagram $H^{(4)}_d$ as function on sublattice, and then we
can apply to it the Fourier transformation rules for the sublattice. First
stage does not differ from the usual perturbative lattice calculations.
Resulting functions are, for example, as
\begin{equation}
F_1(X_1,\; Y_1)=\int_{-\pi/a}^{\pi/a} {\rm e}^{i(X_1-Y_1)k} \dbar(k)G(k).
\end{equation}
This, however is not the case when the remaining Fourier transforms and
summations over internal sublattice points are performed. Due to the difference
between the Brillouin zones result will be sums rather then simply products.
For instance, for the block $F_1$ we will obtain:
\begin{equation}
 F_1(K)= \sum_{all \; n_\mu=1}^\eta \left[\sum_\mu \frac {2}{\eta^{(d-2)}} {\rm
cos}\left(\frac{K_\mu+2 \pi n_\mu}{\eta}\right)\right] \frac {1}{\eta^2
\sum_\mu 4{\rm sin}^2 (\frac{K_\mu+2 \pi n_\mu}{2\eta})}
\end{equation}
Such an expressions we will call "decimated" and denote by $[| "expr" |]$:
 \begin{equation}
[|f(K,L,...)|]\equiv \sum_{(n_{K\mu},n_{L\mu},..)=1}^\eta f(K+2 \pi n_K,L+2 \pi
n_L,...)
\label{decimator}
\end{equation}
Notice that the decimated function does not possess original translation
invariance with the period $2 \pi /a$, but instead period is $2 \pi /(\eta a)$.
This corresponds to the nature of decimation as transformation ${\cal
L}\rightarrow {\cal L}^*$.
Eventually, diagram $G^{(4)}_d$ takes the following form:
\begin{equation}
H^{(4)}_d(K,P, Q)=[|F_1(K)|]{\bf G}^{-1}(K)[|F_1(P)|]{\bf
G}^{-1}(P)[|F_3(K,P,Q,-K-P-Q)|].
\end{equation}

The described algorithm is applicable not only to the tree level contributions.
When the representation of the propagator eq.(\ref{int-prop}) is used, some of
the loop diagrams will contain ${\bf G}^{-1}$. If this is the case, one should
calculate the corresponding decimated functions, and only then perform the loop
integration.

We applied the formalism of the real space RG to $O(N)$ symmetric nonlinear
sigma model. In $d$ dimensions, this model is described by action
\begin{equation}
\label{origaction}
{\cal A}= -\frac {1}{2 g} \sum_x S^a\Box S^a,
\end{equation}
where $S^a$ ($a=1,..,N$) is $O(N)$ vector normalized on unity, $S^2=1$, $g$ is
the coupling constant.
The most general covariant effective action with up to four lattice derivatives
is:
$$
{\cal A}^{eff}=-g^{-1} \sum_X \left[ \frac {1}{2} S^a\Box S^a+c_5(\Box
S^a)^2+(-\frac {1}{24} +c_6) \sum_\mu(\partial_\mu \partial_\mu^+ S^a)^2\right.
$$
\begin{equation}
\label{effaction}
\left.+c_7(S^a\Box S^a)^2+c_8\sum_\mu (S^a \partial_\mu \partial_\mu^+
S^a)^2+c_9\sum_{\mu \nu} \left (\frac {\partial_\mu + \partial_\mu^+}{2} S^a .
\frac {\partial_\mu + \partial_\mu^+}{2} S^a\right)^2 \right ].
\end{equation}
Note that Symanzik-improved action \cite{Symanzik} contains $O(N)$ noncovariant
terms. These can be covariantized \cite{Elitzur} introducing source terms like
$J\Box S$, $J^2$. In real space decimation these obviously do not appear.

Decimation technique is applied to the unconstrained variables, while both
original and effective actions eqs.(\ref{origaction},\ref{effaction}) are
written in terms of constrained, covariant fields $S^a$. To build a
perturbation theory, one can solve the constraint:
$$
S^a=(\sqrt{g} a^{(d-2)/2} \pi^i,\sqrt {1-g a^{(d-2)} (\pi^i)^2}), \;
i=1,..,N-1,
$$
and expand these actions in terms of "pions" $\pi^i$. Then to the fourth order
in $\pi^i$ and up to fourth derivatives, the effective action is:
$$
{\cal A}^{eff}={\cal A}^{(2)}+{\cal A}^{(4)}
$$
with the quadratic part
\begin{equation}
{\cal A}^{(2)}= -\sum_X \left[ \frac {1}{2} \pi^i\Box \pi^i
+c_5(\Box \pi^i)^2+(-\frac {1}{24} +c_6) \sum_\mu(\partial_\mu \partial_\mu^+
\pi^i)^2\right]
\end{equation}
and the quartic part
$$
{\cal A}^{(4)}=-g \sum_X \left[ \frac {1}{8} \pi^2\Box \pi^2
+ \frac {1}{4}c_5(\Box \pi^2)^2+\frac {1}{4}(-\frac {1}{24} +c_6)
\sum_\mu(\partial_\mu \partial_\mu^+ \pi^2)^2 \right.
$$
\begin{equation}
\label{effexpand}
\left. + c_7(\pi^i\Box \pi^i)^2+ c_8\sum_\mu (\pi^i \partial_\mu \partial_\mu^+
\pi^i)^2+ c_9\sum_{\mu \nu} \left (\frac {\partial_\mu + \partial_\mu^+}{2}
\pi^i . \frac {\partial_\nu + \partial_\nu^+}{2} \pi^i \right )^2 \right].
\end{equation}
To reconstruct the coefficients $c_5$,..,$c_9$, we perform the perturbative
decimation of the original theory, obtaining the decimated coefficient
functions $H^{(n)}$. Then the coefficients can be identified by simple
comparison. We applied the method to $d=2$ $\sigma$ model for $\eta =2$.
Comparing the quadratic parts at tree level, we find $c_5,\; c_6$:
\begin{equation}
c_5=5/32; \quad c_6=1/96.
\end{equation}
There are three unknown coefficients in the quartic part of the effective
action eq.(\ref{effexpand}). To determine them, we calculated the quartic
decimated function $H^{(4)}$ for three different momenta configurations, thus
obtaining a system of the linear equations for $c_7,\; c_8,\; c_9$ with the
solution:
\begin{equation}
c_7=c_8=0,\, c_9=-5/64.
\end{equation}
The vanishing of $c_7$ and $c_8$ is rather surprising. We do not see any
obvious reason for this.

To see how formalism works we applied it to the solvable $d=1$ $O(N)$ symmetric
$\sigma$ model at both tree and one loop level. In this case there is only one
quartic term, $\frac {c\; g}{a^2}\sum_X (\pi^i_X \Box \pi^i_X)^2$ in the
effective action.
We calculated two and four point functions.
These perturbative RG results were compared to expansion of exact solution for
$\eta =2$ decimated
 effective action of Heisenberg chain \cite{Rosenstein}:
\begin{equation}
-{\cal A}^{eff}=(1/(2 g)-(N-2)/4) (S^a_X S^a_{X+1}-1)-(1/(16 g)-(N-2)/16)
(S^a_X S^a_{X+1}-1)^2+...
\end{equation}
and are in complete agreement.

We do not present any technique details here. Only note, that at the first
sight the decimated coefficient functions $H^{(2)}$ should behave as $1/P^2$
and $H^{(4)}$ as $1/P^4$. In fact, however, due to the decimation procedure all
the negative powers of momenta and the constant terms cancel. This exact
cancellations provide an additional consistency check.

To summarize we found a systematic way to perform RG of the decimation type in
$d>1$ perturbatively. Here we would like to discuss some of its uses. The
formalism we propose here preserves all the local relations including
constraints. This in turn means that effective (coarse grained) theory will
obey exactly the same local constraints as did original, and that no
non-covariant terms will appear in the effective action.

Another, compared to others \cite{Migdal-Kadanoff,Patkos,Swendsen}, useful
feature of proposed formalism is its perturbative character. This can provide
us with systematic method of calculations in asymptotically free models and,
what is even more essential, with a way to do {\it controllable}
approximations. Hopefully, this side of proposed formalism will make it
applicable in situations when such a control is essential, as in the recently
proposed double strong-weak expansion approach \cite{Rosenstein}. It turns out
that in asymptotically free theories there exist a region in the parameter
space in which both strong and weak coupling expansions are valid at the same
time. Namely both the practical weak coupling $\alpha(g)=const\;g$ and strong
coupling $\beta(g)=const/g$ expansion parameters there are reasonably small.
The "loop
factors" $1/(4\pi)^2$  in the practical weak coupling expansion
parameter $\alpha(g)$ are partly responsible for this. In this scheme high
frequency modes are integrated out perturbatively and the resulting effective
action treated using strong coupling expansion. The symmetry preserving
perturbative decimation technique is the most suitable tool for the first part
of such calculations.

We would like to stress also that the method described here unlike most of the
other decimation (and exact RG in general) techniques, enables us to perform
decimations not restricted to the simplest case $\eta=2$ only. The $\eta >2$
calculation just takes a bit more computer time.

Discussions with B. Hu and H. Miller are greatly appreciated.
This work was supported by
National Science Council of ROC, grants
NSC-83-0208-M-001-011 (B.R.) and NSC-83-0208-M-001-015 (V.K.).

\end{document}